%% file: Mechatronics2014.tex
\documentclass[conference,a4paper]{IEEEtran}
\pdfoutput=1
\newcommand\sub[1]{\ensuremath{\sb{\mathrm{#1}}}}

\usepackage{graphicx}
\usepackage[T1]{fontenc}
\usepackage[ansinew]{inputenc}
\usepackage[english]{babel}
\usepackage{booktabs}
\usepackage{placeins}
\usepackage{tabularx}
\usepackage{tabulary}
\usepackage{multirow}
\usepackage{rotating}
\usepackage{xcolor,colortbl}
\usepackage[tight,nice]{units}
\usepackage{upgreek}
\usepackage{textcomp}
\usepackage{flushend}
\usepackage{gensymb}
\usepackage[caption=false]{subfig}

\usepackage{pdfpages}

\usepackage{needspace}
\graphicspath{{media/}{media/jpg/}{media/pdf/}{mediaold/}{mediaold/jpg/}{mediaold/pdf/}}

\begin{document}

\title{Wireless distance estimation with low-power standard components in wireless sensor nodes}

\author{\IEEEauthorblockN{Thorbjörn Jörger\IEEEauthorrefmark{1}\IEEEauthorrefmark{2},
Fabian Höflinger\IEEEauthorrefmark{1},
Gerd Ulrich Gamm\IEEEauthorrefmark{1} and
Leonhard M. Reindl\IEEEauthorrefmark{1}}
\\
\IEEEauthorblockA{\IEEEauthorrefmark{1}Department of Microsystems Engineering\\
University of Freiburg - IMTEK,
79106 Freiburg, Germany\\ \{thorbjoern.joerger, fabian.hoeflinger, gerd.ulrich.gamm, reindl\}@imtek.de}
}

\maketitle

\begin{abstract}
In the context of increasing use of moving wireless sensor nodes the interest in localizing these nodes in their application environment is strongly rising. For many applications, it is necessary to know the exact position of the nodes in two- or three-dimensional space. Commonly used nodes use state-of-the-art transceivers like the CC430 from \textsl{Texas Instruments} with integrated signal strength measurement for this purpose. This has the disadvantage, that the signal strength measurement is strongly dependent on the orientation of the node through the antennas inhomogeneous radiation pattern as well as it has a small accuracy on long ranges. Also, the nodes overall attenuation and output power has to be calibrated and interference and multipath effects appear in closed environments.\\
Another possibility to trilaterate the position of a sensor node is the time of flight measurement. This has the advantage, that the position can also be estimated on long ranges, where signal strength methods give only poor accuracy. In this paper we present an investigation of the suitability of the state-of-the-art transceiver CC430 for a system based on time of flight methods and give an overview of the optimal settings under various circumstances for the in-field application. For this investigation, the systematic and statistical errors in the time of flight measurements with the CC430 have been investigated under a multitude of parameters. Our basic system does not use any additional components but only the given standard hardware, which can be found on the \textsl{Texas Instruments} evaluation board for a CC430. Thus, it can be implemented on already existent sensor node networks by a simple software upgrade.\\

\let\thefootnote\relax\footnotetext{\IEEEauthorrefmark{2}Corresponding author}

\end{abstract}


%
\IEEEpeerreviewmaketitle

\section{Introduction}
There exist many solutions for wireless distance measurement and equipment specifically designed for this purpose. Our approach is different as we implement precise wireless distance measurement in already existing and used standard hardware. A lot of available devices nowadays have an integrated wireless transceiver for communication purposes on board. We use these existing wireless transceivers for localizing the devices. There is a huge demand for information on the whereabouts of devices related to their environment. A possible application is the navigation of pedestrians in railway stations, airports, trade fairs or department stores. Additionally autonomous robots can be guided in processing plants without the need for a human operator. A lot of the localization tasks must be realized in indoor environments where no GPS signal is available or GPS receivers have too high energy consumption. To provide a cheap and low power localization, the time of flight measurement (ToF) of a transmitted signal can be used to trilaterate the position of an object in three-dimensional space. With the combination of signal strength and runtime measurement, the accuracy and overall reliability of the system can be further enhanced.\\
The key issue in runtime measurements with the most wireless sensor nodes is the relatively slow system clock. It generates an uncertainty of one clock cycle which is negligible in ultra sonic sensing due to the slow speed of sound but becomes overwhelmingly large when working with electromagnetic waves which are traveling with the speed of light. With a \unit[26]{MHz} counter clock one clock cycle lasts \unit[38.4]{ns}. Translated to distance this clock jitter is equivalent to an error of approximately \unit[11.5]{m}. Additionally the hardware generates a time offset of several hundred cycles or around \unit[6-8]{\textmu s} between physically receiving a packet and logically rising the respective \mbox{RX/TX} flag which triggers the integrated counter capturing. This offset in the analog domain of the transceiver is moreover dependent on the attenuation, the frequency and the chosen modulation. We researched influences of the aforementioned parameters on the delay and the overall performance in respect to distance measurement and present settings which promise to be the best foundation for further research in this field.\\
The paper starts with an overview of related work in Section~\ref{sec:related}. In Section~\ref{sec:theory} the theoretical background of our research is presented and the used equations are given. The programming of the hardware is explained in Section~\ref{sec:programming}. Section~\ref{sec:methods} presents the methods used in the experimental part of our work. In Section~\ref{sec:discussion} the results of the measurements are presented and subsequently discussed. Finally, Section~\ref{sec:conclusion} gives a conclusion of the research done in the paper and an outlook for research in the future.
\section{Related work}
\label{sec:related}
For localization of low cost sensor nodes, most of the existing applications and research papers use the signal strength indicator~\cite{paperindoor}. The behavior of radio signals in an indoor environment was shown by Hashemi in~\cite{articlehashemi}. Via the received signal strength indicator (RSSI) the distance between two wireless transceivers can be calculated. In most cases the two transceivers will consist of an anchor node and a mobile tag.\\
Several groups have researched combinations between round trip time (RTT) and RSSI measurements, but for Wi-Fi applications~\cite{paperRSSRTT, paperwifilocation, paperdistanceRTS} and considerations of Cramér-Rao bounds~\cite{paperCramerRaoTOA, paperCramerRaoLower}. In~\cite{paperberlin} Will et al. have analyzed a multitude of error mechanisms and influences on their ToF system, which also measures the RTT. They use the mean value of 25 single measurements to find the distance between two nodes. We try to overcome the physical limitations of single time measurements with more extensive statistical methods and a precise characterization of the underlying error mechanisms. Also we take into account the different behavior in respect to attenuation, modulation and data rate.\\
The wake-up strategies mentioned in the abstract can be used to lower the overall energy consumption. Wireless sensor nodes based on the CC430 are capable of utilizing low power modes, in which they consume only power in the range of \unit[several]{\textmu W} compared to about \unit[100]{mW} during radio operation~\cite{gamm2012low, vanderdoorn09}. This allows the extension of battery life and operational lifetime by several orders of magnitude~\cite{Wendt08IMTC}. A view on cooperative localization in wireless sensor networks and the underlying algorithms can be found in~\cite{paperpatwari}. Wendeberg et. al. provide spring-mass based self-localizing evaluation algorithms in~\cite{wendeberg2013calibration}, which can be fed by the data generated by this system.

\section{Theoretical background}
\label{sec:theory}
\subsection{Received signal strength}
The following Friis equation calculates the maximum possible power in a receiving node $P\sub{r}$ dependent on the sending power $P\sub{t}$, the antenna gains $G\sub{r}$ and $G\sub{t}$, the distance $r$ and the wavelength $\lambda$: 
\begin{equation}
	{P\sub{r}}={P\sub{t}}\cdot{G\sub{t}}\cdot{G\sub{r}}\cdot\left (\frac{\lambda}{4 \, \pi \, r}\right)^2.
	\;
	\label{eq:Freiraumdaempfung}
\end{equation}
Therefore, if the received power strength is measured in the receiving node, the equivalent distance of sender and receiver can be calculated with: 
\begin{equation}
	r=\frac{\lambda}{4\,\pi}\cdot\sqrt{\frac{P\sub{t}}{P\sub{r}}\cdot G\sub{t}\cdot G\sub{r}}.
	\;
	\label{eq:Freiraumd}
\end{equation}
The accuracy depends on the number of anchor nodes and of the environment, which can lead to dynamic errors. The RSSI value is distorted by objects in the direct path, in the vicinity and by environmental influences, especially air humidity. Additionally, the RSSI value also depends on the orientation of the antenna. The antenna directivity is influenced by the antenna type and the spatial orientation of it compared to the anchor nodes. Using an RF system for localization, people can be localized with low accuracy (\unit[1.5-3]{m}). Through combination with other technologies, this accuracy can be improved. The multi-method approach in~\cite{wendeberg2011} uses a combination of built-in sensors of mobile devices and the capabilities of the end-users, which estimates positions with a scanner application.

\subsection{Time of flight}

\begin{figure}
\centering
\includegraphics[width=3.5in]{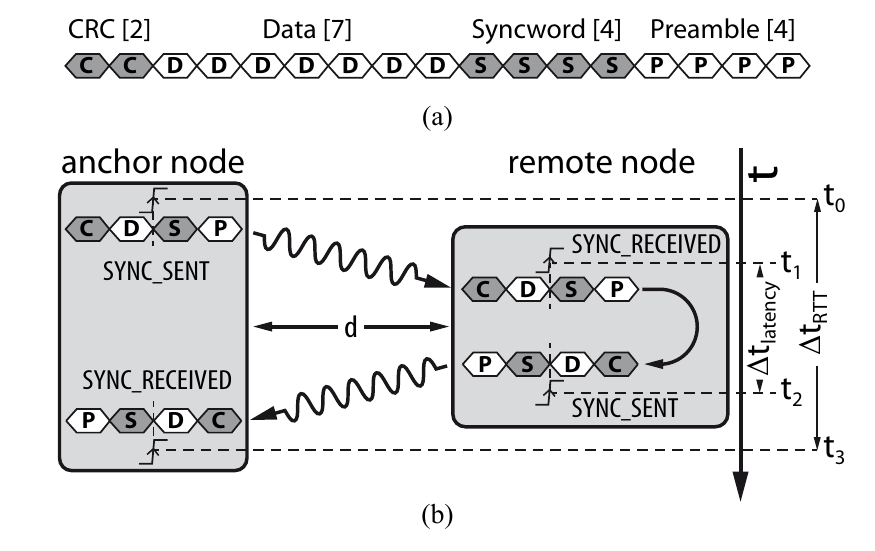}\vspace{-5pt}
\caption{a) Packet structure and size. The numbers in brackets stand for the size in bytes corresponding to the number of byte hexagons. b) Schematic of the timing parameters of the measurement setup. The time indices mark the rising flags on which each counter capture is triggered by an incoming or outgoing sync word. The figure depicts one round trip. This same measurement is repeated for several hundred times in a row.}
\label{fig:setup}
\end{figure}

Our approach uses the time of flight (ToF) measurement to estimate the distance between the anchor node and the mobile tag and to overcome the above mentioned disadvantages of RSSI measurements. We use a special form of ToF that is called round trip time as described in Figure~\ref{fig:setup}. Compared to the normal ToF measurement the signal of RTT travels the way back and forth from the anchor node to the mobile tag. Therefore no synchronization between anchor node and mobile tag is needed. The anchor node calculates the time difference. The mobile tag is only measuring the processing latency between incoming signal and outgoing signal. 
By measuring in the anchor node the moment of sending the package $t\sub{0}$ and the moment of receiving back the package $t\sub{3}$, the complete roundtrip time can be calculated with $\Delta t\sub{RTT}=t\sub{3} - t\sub{0}$ including the latency. By measuring in the remote node the moment of receiving the package $t\sub{1}$ and the moment of sending the package $t\sub{2}$, the additional signal processing latency $\Delta t\sub{latency}=t\sub{2}-t\sub{1}$ in the mobile tag can be calculated. The corrected round trip time $t\sub{RTT,corr}$ is then 

\begin{equation}
	\Delta t\sub{RTT,corr}=\Delta t\sub{RTT} - \Delta t\sub{latency}.
	\;
	\label{eq:Entfernung_RTT1}
\end{equation}
This round trip time is in our experiments between \unit[6]{\textmu s} and \unit[1400]{\textmu s} depending primarily on the chosen modulation and data rate. It is therefore assumed, that 
\begin{equation}
	t\sub{signal}=\Delta t\sub{RTT,corr}-t\sub{offset}
	\;
	\label{eq:offset}
\end{equation}
with a, for each setting, constant offset in the analog domain and the true signal runtime delay $t\sub{signal}$. The distance $d$ can then be calculated according to
\begin{equation}
	d=\frac{ v  \cdot t\sub{signal} }{2}
	\;
	\label{eq:Entfernung_RTT2}
\end{equation}
with $v=c \cdot v\sub{p}$ where $c$ is the speed of light and $v\sub{p}$ the velocity factor of the medium, which can be either air or the copper of the coaxial cable.

\subsection{Processor cycles}
The first versions of our system used the highest clock rate of the CC430 of \unit[20]{MHz} for the time measurement. This proved to be extremely error prone due to the fact, that the internal RC oscillator of the CC430 is dependent on operating voltage with \unit[1.9]{\%/V} and ambient temperature with \unit[0.1]{\%/\celsius}. With a mean round trip time of then about 30000 cycles and \unit[30]{mV} operating voltage difference, the distance error increased to unusable \unit[270]{m}. Fortunately, the CC430 has the ability, to drive the counter with an external clock, and with the \unit[26]{MHz} crystal oscillator onboard the used evaluation boards, such an external clock was already available. The onboard crystal oscillator provides an overall accuracy of \unit[80]{ppm} which leads to a clock generated distance error of a single measurement of \unit[3.2]{m}. An accurate crystal with small temperature and operating voltage dependency is crucial for precise distance measurements.

\subsection{Modulation}
Binary frequency shift keying \mbox{(2-FSK)} is a modulation, where the information is transferred by shifting the carrier frequency by a certain amount in one direction or the other. It is one of the simplest digital modulations and widely used in wireless transceivers. Gaussian shaped frequency shift keying \mbox{(2-GFSK)} differs from standard \mbox{2-FSK} in the fact that a Gaussian filter is applied to the output signal which smoothes the pulse and so reduces the spectral bandwidth of the signal.\\
The influence of modulation on the aforementioned time delay in the analog domain is dominant. The time delay of \mbox{2-GFSK} modulation is around \unit[40]{\%} higher at every data rate than the time delay of the \mbox{2-FSK} modulation. Apparently, the processing of the \mbox{2-GFSK} signal takes up more time in the analog domain, then the processing of the variant without the application of a Gaussian filter. Additionally, the standard deviation of the single measurements doubles, when a Gaussian filter is used. Hence we recommend the use of \mbox{2-FSK} for all distance measurement purposes when feasible.

\subsection{Frequency}
Fresnel described the implications of objects being in the so called Fresnel zone $F\sub{n}$. The radius of the $n$-th Fresnel zone at the point $P$ is derived with the wavelength $\lambda$, the distance between antennas $d$ and the distances $d\sub{1}$ and $d\sub{2}$ of the antennas to the respective point $P$ according to
\begin{equation}
	F\sub{n}=\sqrt{\frac{n\,\lambda\,d\sub{1}\,d\sub{2}}{d}}.
	\;
	\label{eq:fresnel}
\end{equation}
Obstacles in the uneven Fresnel zones reflect the incident wave in a way, that destructive interference occurs and the attenuation rises. To get a small attenuation in radio communication the first Fresnel zone $F\sub{1}$ has to be clear of obstacles, which mostly cannot be ensured in indoor environments. Interference patterns in closed environments arise by multipath wave propagation and can lead to situations where no communication is possible because reflected waves cancel each other completely out. Because the interference pattern is, according to Bragg's law, directly dependent from an integer multiple $n$ of the wavelength $\lambda$ 
\begin{equation}
	n \lambda = 2 d \sin \theta,
	\;
	\label{eq:bragg}
\end{equation}
a possible method to circumvent such a scenario would be to change the working frequency whenever a node is encountered with a situation where no connection to the anchor node is possible. \\
For this work, two commonly used frequencies, \unit[868]{MHz} and \unit[915]{MHz}, were used. The difference in frequency would be sufficiently large to change the interference pattern in a way that a connection to the anchor node could be reestablished. 
\begin{figure}
\centering
\includegraphics[width=3.5in]{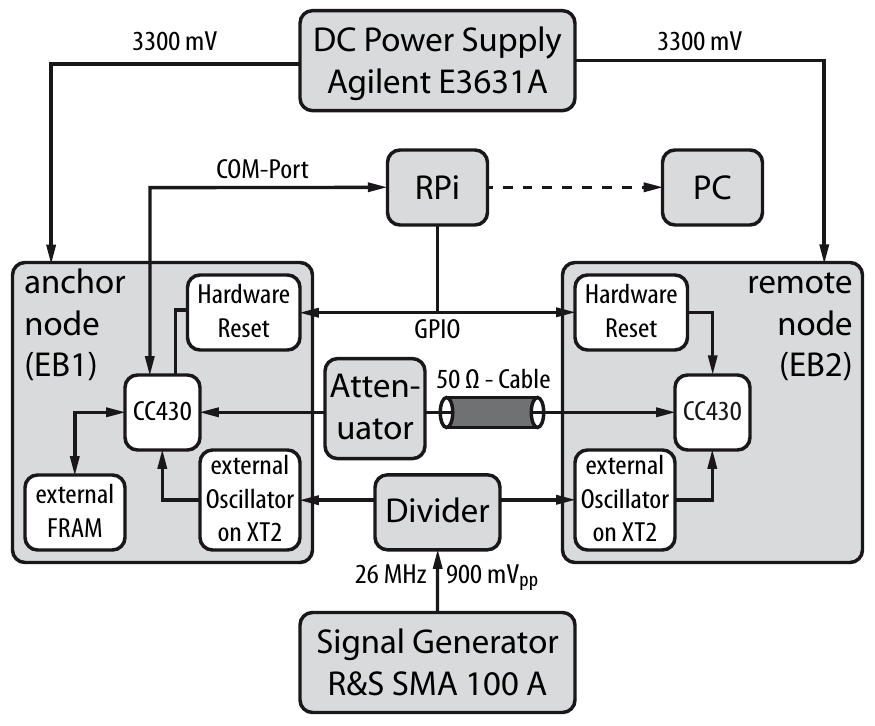}
\caption{Block diagram of the system setup. The two evaluation boards are connected with a coaxial cable. The Raspberry Pi (RPi, see Section~\ref{sec:sub:RPi}) starts the measurement by sending a start command containing the setting and the number of single measurements to perform to the anchor node. The anchor node (master) is then initiating the communication with the remote node (slave) and stores all incoming and measured data into the \unit[2]{MBit} FRAM. After finishing the scheduled amount of measurements, the stored data are transferred to the RPi via COM port.}
\label{fig:syssetup}
\end{figure}
\subsection{Time delay in the analog domain}
One of the most interesting effects encountered, is the time delay in the analog domain. It is not documented in the datasheets and manuals from \textsl{Texas Instruments}. It is speculated, that the delay depends on the internal programming of the state machine, the rise times of the preamplifiers and the data processing of the sync word. The delay seems to stay constant over time and is in the same range between different evaluation boards, but preliminary results show, it is temperature and supply voltage dependent to some degree. It changes strongly between frequencies, data rates and modulations. Further investigation of this effect is necessary to cancel out its effects in the final application.
\section{Programming}
\label{sec:programming}
\subsection{State machine}
For the time measurement, a finite state machine has been implemented in both evaluation boards. The programming is mostly identical. The only differences exist in the state machine, where the master is allowed to initiate the measurement procedure while the slave is not, as well as the interrupt service routine, where the master counts the time between sending and receiving while the slave counts the time between receiving and sending a packet. To start the measurement procedure, a data package has to be received via the COM port. It contains the necessary information about the number of single measurements that should be taken and the RF settings that should be used. The master then sends this information to the slave, which adjusts its configuration accordingly. After reestablishing the communication with the new configuration, the scheduled number of measurements is taken. During this time, the master is not communicating with the Raspberry Pi (RPi, see Section~\ref{sec:sub:RPi}) nor accepting any input from it in order to take a clean and undisturbed measurement. Therefore the data transfer to the RPi is initiated by the master.
\subsection{Data transfer}
The measurement data is recorded by the evaluation boards designated as master. It measures its own time span and received signal strength and receives the time span and the RSSI from the remotely installed second evaluation board designated as slave. The data from the remote evaluation board is simply piggy-backed onto the packages sent and received anyway. All information is then sent to the RPi via a standard COM port. The RPi administrates the data reorganization and storage as well as all calculations.
\subsection{Time measurement}
\begin{figure}
\centering
\includegraphics[width=3.5in]{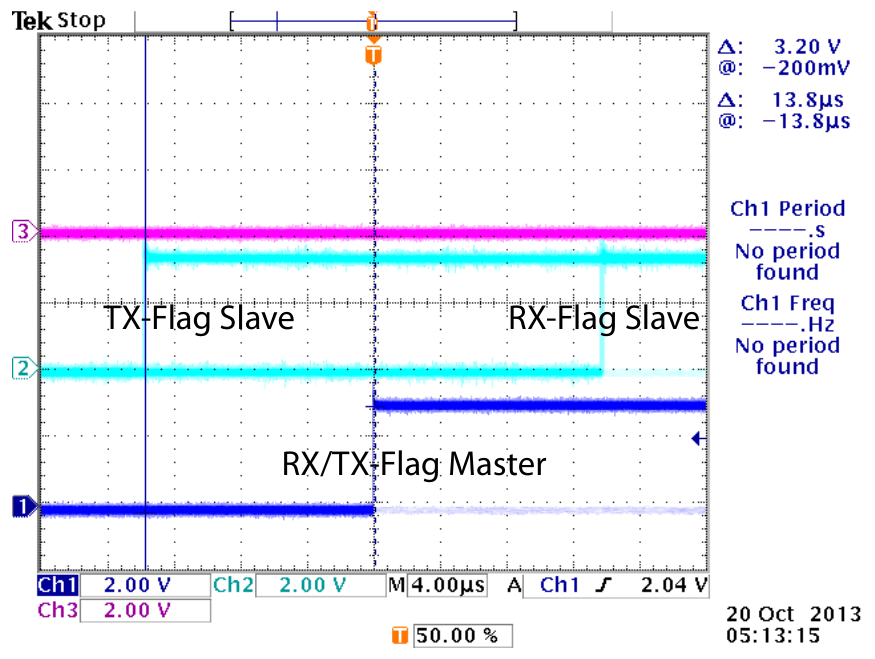}
\caption{Time difference between rising of the sending flag (TX) on one node and rising of the receiving flag (RX) of the same packet on the other node is much larger than it is supposed to be by the runtime effect only. The runtime is superimposed by the time delay in the analog domain. The oscilloscope was connected to the GDO1s of the CC430s with pin setting: IOCFG1 = 0x06. The cable length between both evaluation boards was \unit[2]{m}.}
\label{fig:rxtx}
\end{figure}
In order to achieve a precise time measurement without additional hardware, the internal 16-bit counter of the CC430 was used. It was fed by the external 26-MHz crystal RF oscillator already available on the evaluation board to use the highest allowed and available frequency. The corresponding period is \unit[38.4]{ns} and as such the smallest resolvable time unit in this system, further referred to as cycle.\\
The largest round trip time resolvable is $2^{16}$ cycles or approximately \unit[2.5]{ms}, which was not enough for our purpose. Hence the counter was enlarged with an additional 16-bit register and an overflow interrupt service routine to provide $2^{32}$ cycles or approximately \unit[165]{s} capacity.\\
The counter is triggered by the rising internal sync word received/sent flag. The counter value is stored in the corresponding register and then saved by an interrupt service routine for calculating the time difference between the occurrences of the triggered events. This measurement is repeated for at least several hundred times. Due to this many samples, a very precise measurement value can be obtained.\\
To ensure the reliability of the measured events, the values were compared to measurements taken with a \textsl{Tektronix} TDS3034B oscilloscope. Figure~\ref{fig:rxtx} shows the oscilloscope analysis of the micro controller flags utilized for the time measurement. This was used to confirm the validity of the time measurements with the integrated counter of the CC430. All values were inside the measurement resolution of the oscilloscope and therefore this is an apt method of time measurement for our application.
\section{Material and method}
\label{sec:methods}
For the sake of reproducibility and conformability, all measurements were performed with hardware freely and commercially available. The system setup used for all measurements is depicted in Figure~\ref{fig:syssetup}. In this section, you will find a listing of all necessary hardware and the measurement methods used.
\subsection{Statistical evaluation}
For statistical evaluations of measurements, the standard deviation estimation correction factor
\begin{equation}
   c\sub{\sigma} = \sqrt{\frac{N-1}{2}}\cdot\frac{\Gamma\left(\frac{N-1}{2}\right)}{\Gamma\left(\frac{N}{2}\right)},
   \;
\label{eqn:stdestimator}
\end{equation}
is used to calculate the influence of small numbers of measurements to a populations standard deviation. It is for the smallest amount of measurements taken for this paper of \unit[10]{k} measurements around \unit[25]{ppm}. This means that the number of samples is sufficient to estimate the standard deviation with an error smaller than \unit[25]{ppm}.\\
For a given population mean value with a certain population standard deviation $\sigma\sub{1}$, the resulting standard deviations of the means $\sigma_N$ over a batch of $N$ measurements is calculated with
\begin{equation}
   \sigma_N = \sigma\sub{1}\cdot\sqrt{\frac{1}{N}}.
   \;
\label{eqn:errorofmeans1}
\end{equation}
To get the number of measurements it takes to get a mean value of the batch near the population mean with a desired uncertainty $\sigma\sub{x}$, Equation~\ref{eqn:errorofmeans1} could be written as
\begin{equation}
   N = \frac{\sigma\sub{1}^2}{\sigma\sub{x}^2}.
   \;
\label{eqn:errorofmeans2}
\end{equation}
\begin{figure}[htb]
\centering
\includegraphics[width=8cm]{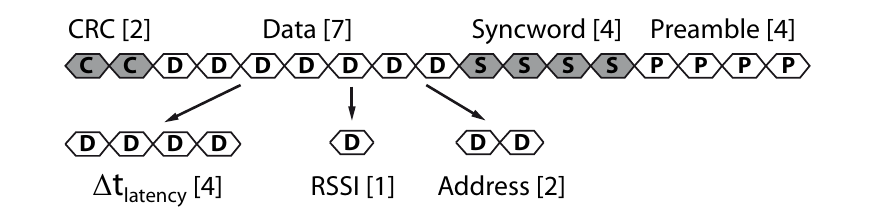}
\caption[Packet structure, content and size]{Packet structure and size as well as packet content. The numbers
in brackets stand for the size in bytes. The CRC is calculated for each packet
automatically by the CC430. The sync word is constant and has to be identical
on master and slave. The preamble is generated by the transceivers and used
for PLL lock-in.}
\label{fig:rtt6}
\end{figure}
With those two equations only a good population mean value with a known population standard 
deviation is necessary to calculate the number of measurements needed.\vspace{5pt}
\subsection{Packet structure and size}
The packet size for the experiments conducted in this paper
is 17~byte. The packet consists of a 4~byte long preamble, a
4~byte long sync word, 7~data bytes and a 2~byte CRC check
sum, as depicted in Figure~\ref{fig:rtt6}. The data bytes contain the~4 byte
timer value $\Delta t\sub{latency}$, the~1 byte long RSSI value and a~2 byte
long address for node identification. Because the accuracy of
the system depends on a high number of measurements, it is
desirable to use very short packages and high data rates, so
that more measurements can be conducted in a given time
frame. On the other hand, high data rates require a small
overall attenuation, which is difficult to achieve when distances
increase and the output power is limited.\vspace{5pt}
\subsection{TI evaluation board}
The \textsl{Texas Instruments} EM-CC430F6137-900 evaluation board is a ready-made platform for the single-chip RF-MCU CC430. It can directly be used for the measurements presented in this publication. The MCU supports a frequency range from \unit[433 - 915]{MHz} with four different modulations of which \mbox{2-GFSK} and \mbox{2-FSK} were used in this work while on-off keying (OOK) and minimum shift keying (MSK) were disregarded, because of their rare use in wireless sensor node networks. Also it provides several low-power modes and an overall low power consumption which makes it popular and widely used for low-power energy self-sufficient systems and wireless sensor nodes.\\
The used boards underwent only minor hardware changes, namely soldering connectors on several port pin drill holes and ground taps, bridging the reset button with a NMOS for external reset and replacing the onboard RF crystal oscillator with a tap to feed the MCU with a clock reference from a signal generator. No other modifications were necessary in order to measure the distance.\\
The removed \unit[26]{MHz} quarz on the evaluation boards was replaced with a \unit[26]{MHz} \unit[900]{mV\raisebox{-.4ex}{\scriptsize pp}} feed from a \textsl{Rohde \& Schwarz} SMA100A signal generator in order to provide temperature independent frequency stability and reproducible measurement results as well as a synchronous clock on both evaluation boards. The operating voltage for the evaluation boards of \unit[3300]{mV} was supplied by an \textsl{Agilent} E3631A DC power supply.
\subsection{Raspberry Pi}
\label{sec:sub:RPi}
In consideration of the long uptimes and high reliability, versatility and expandability requirements of the measurement setup, a Raspberry Pi was chosen to serve as a base. It runs Raspbian Wheezy, a debian linux derivative. The RPi communicates with one of the evaluation boards, the anchor node, over a native COM port. It controls every action of the measurement setup, the communication and the data storage and evaluation.
\subsection{RF settings}
Different sets of settings were used to determine the optimal parameters, such as frequency, data rate and modulation, for wireless distance measurement with CC430 based systems. Table~\ref{tab:stdev} contains an overview over the evaluated parameters. They span the whole range of possible settings of the CC430, with data rates between the highest possible setting \unit[250]{kb/s} and the lowest possible setting \unit[1.2]{kb/s}. 
\subsection{Cable connection}
For the distance variation measurements, \textsl{elspec} LL335 low loss high frequency cables of certain lengths and an attenuation of \unit[14]{dB/100\,m} were used to eliminate the influence of multipath effects, reflections and antenna orientation. Several cables of \unit[10]{m} and \unit[20]{m} length were available. To counteract the cable attenuation which varies with its length and to vary the overall attenuation, an array of one hp8494B attenuator and one hp8496B attenuator was used. The connection from and to the attenuator and to the \textsl{elspec} cables was realized with standard RG223 cable of \unit[1]{m} length available in the lab. 
\section{Measurements}
\label{sec:measurements}
To generate an accurate model, both main influences on the time measurement signal had to be investigated, namely distance and attenuation. For each point in the measurements presented 25 rounds with 1000 measurements each were taken. The results were then cleaned of spikes. Every result for which the runtime was negative or greater than \unit[42000]{cycles} was viewed as flawed and removed.  For the RSSI measurements, the first 60 measurements of each 1000-measurement round were discarded, as well as every measurement which deviated more than \unit[$±$3]{dB} or about at least \unit[4]{$\upsigma$} from the mean value. This was necessary since the RSSI register in the CC430 did not provide any plausible values for the first 60 measurements taken and sometimes during the measurement gave single values that were up to \unit[$±$20]{dB} off compared to the mean value. For the remaining values the means and standard deviations were calculated.\\
\subsection{Distance variation}
The distance was varied by varying the length of the cables connecting the nodes. The length of 
the cables was set to \unit[2]{m} and \unit[13-43]{m} in \unit[10]{m} steps. The measurements were 
performed over a period of several days and in an environment without controlled ambient temperature. 
The attenuation was fixed to \unit[-60]{dBm}. The offset value for \unit[2]{m} was subtracted from the 
results of all other distances to normalize the results.
\vspace{18pt}
\subsection{Attenuation variation}
The second main influence is the attenuation. The evaluation boards were working between \unit[-36]{dBm} to 
\unit[-81]{dBm} attenuation. The attenuation was varied in this range in \unit[1]{dB} steps. The measurements 
were performed inside a climate chamber to exclude thermal influence. The temperature inside the chamber was 
set to \unit[20]{\celsius}.\vspace{18pt}
\subsection{Batch size variation}
For the determination of the necessary batch size, a combination of the aforementioned measurements, namely 
\unit[-60]{dBm} attenuation and \unit[18]{m} cable length, was used. For the fast data rates of \unit[250]{kb/s} 
and \unit[38.4]{kb/s} 300k single measurements and 50k single measurements for the slow data rate of \unit[1.2]{kb/s} 
were performed to achieve a stable population mean and standard deviation. These single measurements were then 
grouped to batches of logarithmic sizes from 1 to 5000 samples per batch. The standard deviation of the means 
of the single batches was subsequently calculated and the resulting values translated to distance with the signal 
velocity $v\sub{s}$ of $\unit[9.2244]{m/cycle}$ derived from Equation~\ref{eq:Entfernung_RTT2} with the velocity 
factor of the cable $v\sub{p} = 0.8$.
\begin{figure}\vspace{1pt}
\centering
\includegraphics[width=3.5in]{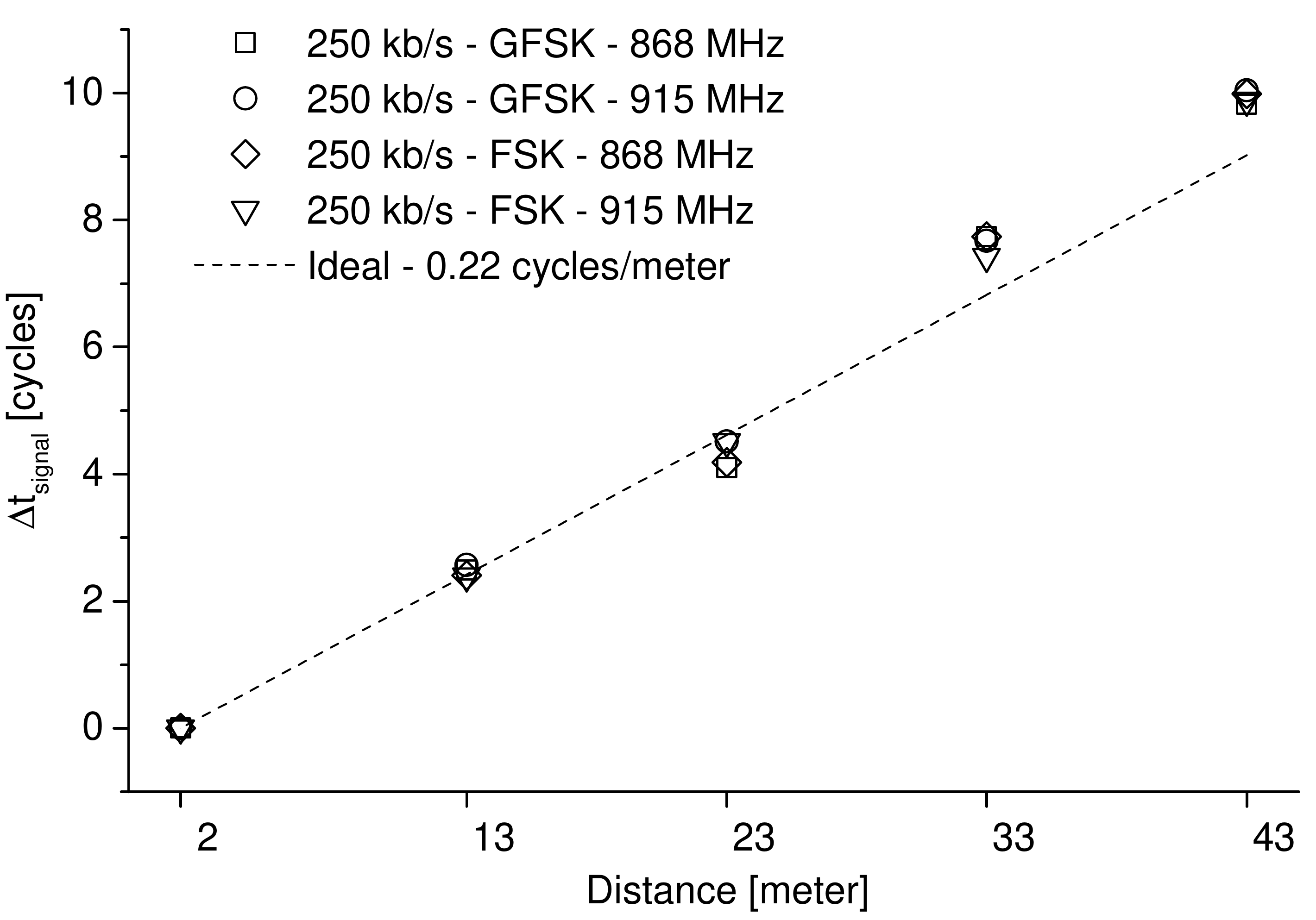}
\caption{Results of reference measurement. The ideal line has a slope of $\unitfrac[0.22]{cycles}{m}$ which corresponds to a velocity factor in the cable of 0.8. The points standard deviation of the means is smaller than the symbol size and was therefore neglected for this graph.}
\label{fig:resdist}
\end{figure}
\begin{figure}
\centering
\includegraphics[width=3.5in]{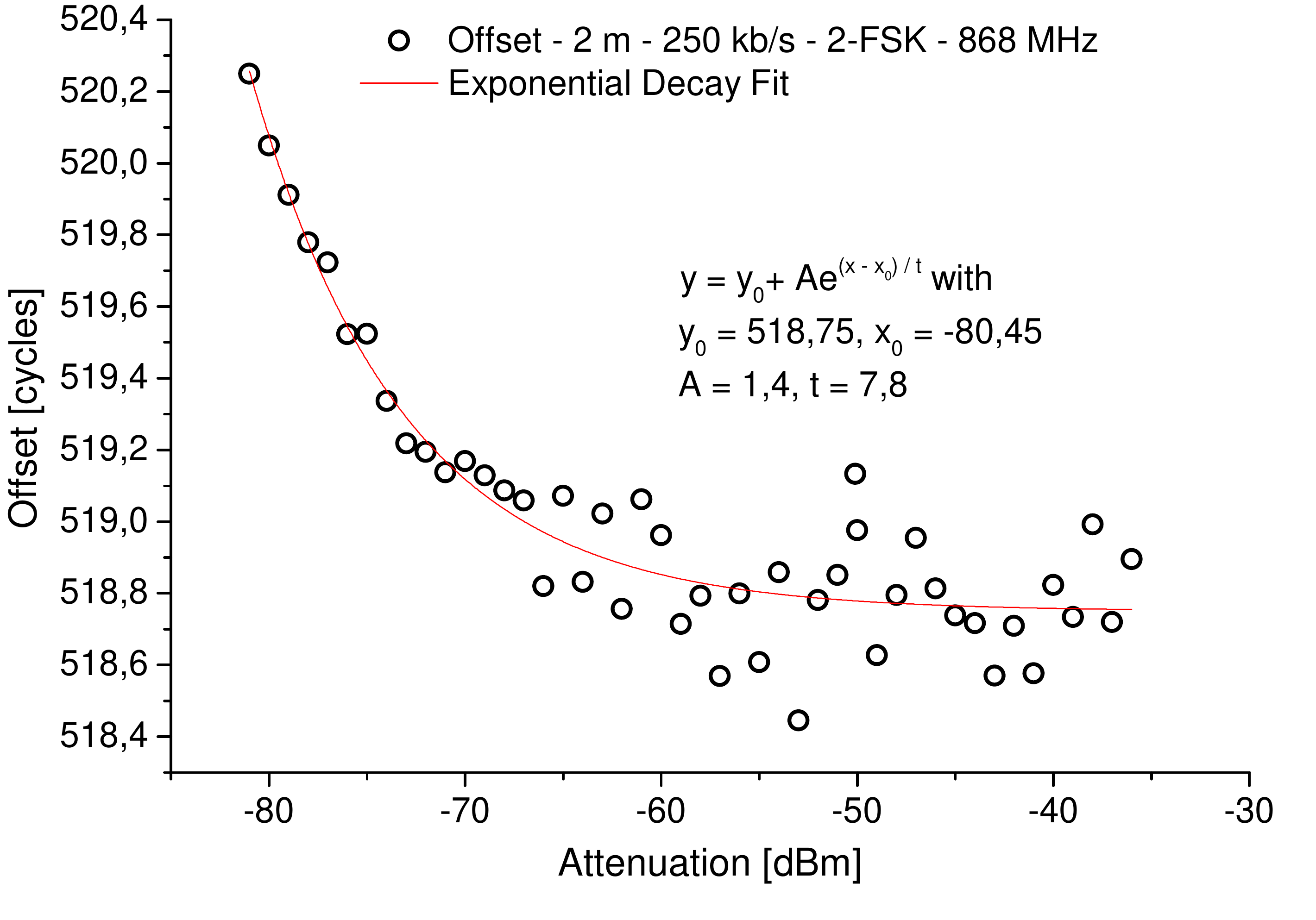}\vspace{-8pt}
\caption{Results of attenuation measurement with a data rate of \unit[250]{kb/s}, FSK modulation and with a frequency of \unit[868]{MHz} in climate chamber at \unit[20]{\celsius}. The exponential decay's fit function is given in the graph.}
\label{fig:atten}
\end{figure}
\begin{figure}
\centering
\subfloat[]{\label{fig:reserror1}\includegraphics[width=3.5in]{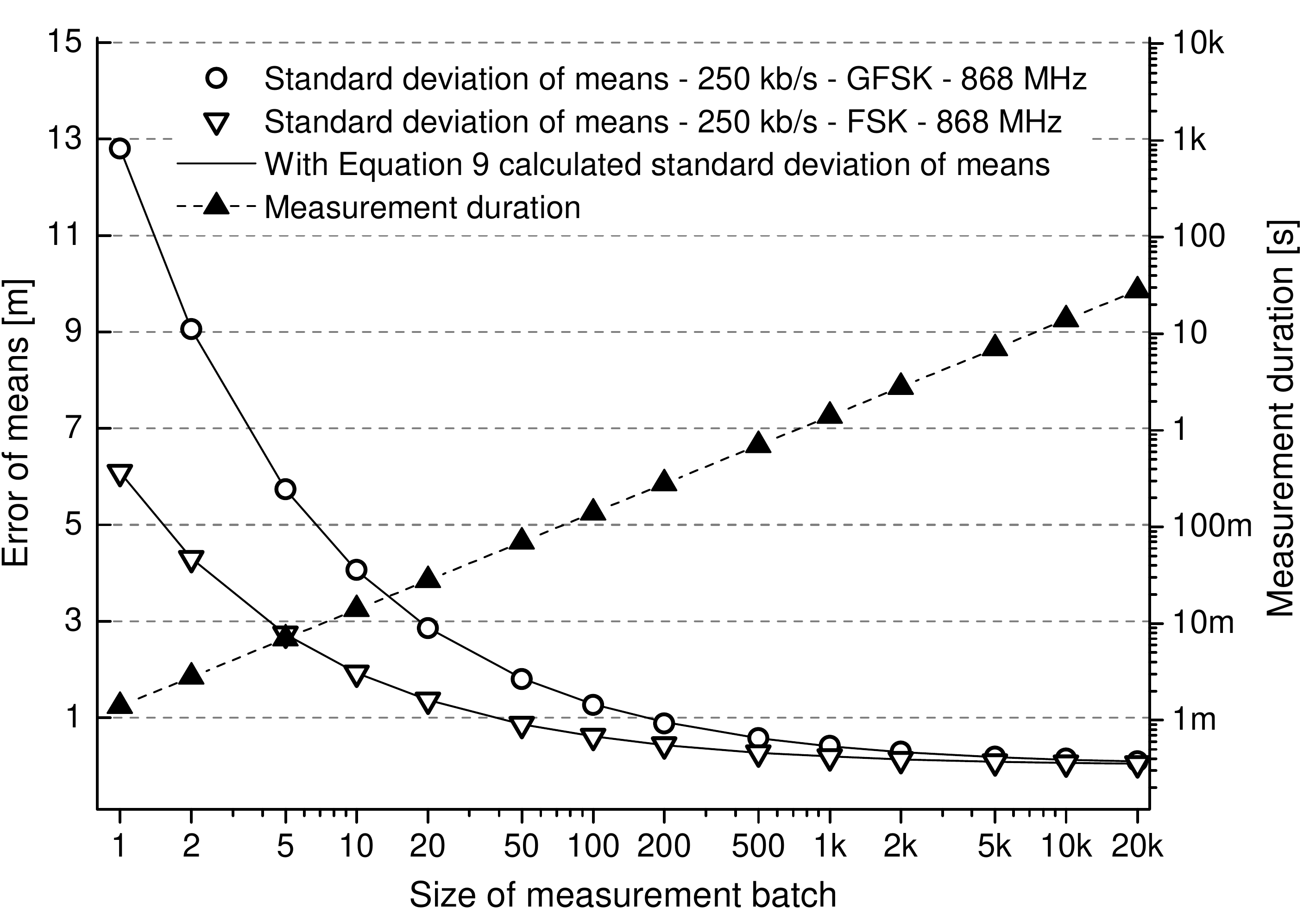}}\\\vspace{-15pt}
\subfloat[]{\label{fig:reserror2}\includegraphics[width=3.5in]{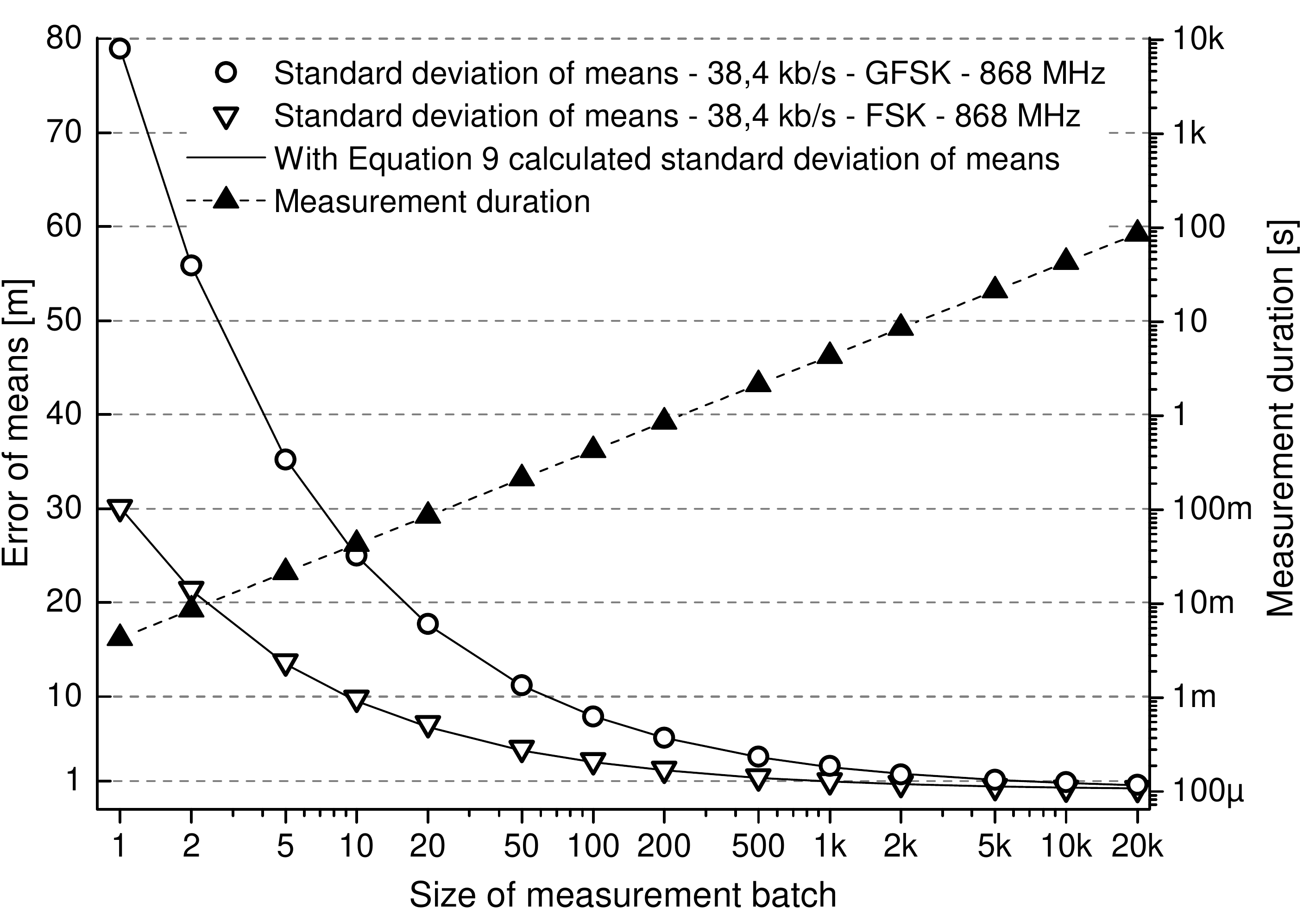}}
\caption{Influence of the batch size to the standard deviation of means as presented in Table~\ref{tab:stdev} for a) \unit[250]{kb/s} and b) \unit[38.4]{kb/s} at a frequency of \unit[868]{MHz} with respect to modulation and the measurement duration. The measured values match the lines calculated with Equation~\ref{eqn:errorofmeans1} from the standard deviation of the population.}\vspace{-10pt}
\label{fig:reserror}	
\end{figure}

\input{media/tablestdevCOLORROTnoRFnoFOOTNOTES}

\section{Results and Discussion}\label{sec:discussion}
The results of the distance measurements as depicted in Figure~\ref{fig:resdist} show the dependence of the round trip time from the distance between the nodes as described by Equation~\ref{eq:Entfernung_RTT2}. The deviations from the ideal line are most probably caused by temperature changes of the environment, because they are close together for all setting. This makes a systematical error more probable. The offset value $t\sub{offset}$  at \unit[2]{m} distance was used to normalize the measurements. After the removal of the offset, only the contribution of the runtime to the whole signal is left.\\
The results of the attenuation measurement is given in Figure~\ref{fig:atten}. The offset increases exponentially with increasing attenuation and therefore with increasing distance. This means, that the measured value overestimates the factual distance when using a linear equation like Equation~\ref{eq:Entfernung_RTT2}. This effect has to be taken into account for future measurements. We suspect the offsets behaviour to be as well temperature dependent so a precise investigation of the offsets behavior is necessary.\\
The results of the variation of the batch size are presented in Table~\ref{tab:stdev} as well as in Figure~\ref{fig:reserror}. It shows the as well the influence of data rate and modulation. The Gaussian filtering of the signal approximately doubles the standard deviation of the measurements. A slow data rate increases the offset and the ratio between offset and desired signal.\\
The time measurement itself is sufficiently precise when  the RF settings and the number of samples are chosen accordingly. This scalability allows a trade-off between speed, accuracy and energy requirements. The number of measurements can as well be calculated with Equation~\ref{eqn:errorofmeans2}.

\section{Conclusion and outlook}
\label{sec:conclusion}
In the paper, wireless distance estimation with the CC430 was evaluated in respect to accuracy and the influence of attenuation. The achievable accuracy for the round trip measurement, when the offset value is measured precisely, is in the decimeter range. The accuracy and speed of the distance measurement with the CC430 are very promising. The measurements can be conducted very fast. A trilateration from four fixed nodes with an accuracy under \unit[0.5]{m} could be performed in one second, while with an accuracy requirement of \unit[1]{m} only \unit[300]{ms} suffice.\\
The second prime target of this experiment was to single out reliable settings for the CC430 to conduct distance measurements. The result clearly show, that \mbox{2-FSK} modulation is superior to \mbox{2-GFSK} modulation by a factor of 2. We conclude, that further research in this field should concentrate on \mbox{2-FSK} modulation, as we will do in the future. With respect to data rate, the result is not as obvious, but instead a trade-off between range, speed and accuracy requirements. We discard the settings with \unit[1.2]{kb/s} data rate, because they are obviously unusable for our objective. We suggest either a system which uses one high and one low data rate and changes between them when the nodes gets out of range, or a system which uses a compromise such as a data rate of \unit[125]{kb/s}.\\
The transmission frequency has no practically relevant influence and can be chosen according to regulations or compatibility to existing systems.\\
Preliminary measurements showed the offset to be temperature and supply voltage dependent. Further research has to be done with respect to the behavior of the offset in the analog domain.  If it cannot be calibrated by precise measurements and modeling, some kind of on-the-fly calibration has to be implemented. We will conduct further research in a climate chamber and under outside free field conditions and report these findings in a future publication.

\section*{Acknowledgment}
This work has partly been supported by the German Research Foundation (DFG) within the Research Training Group~1103 (Embedded Microsystems).

\bibliographystyle{IEEEtran}
\bibliography{MSCref}

\end{document}

%% file: media/tablestdevCOLORROTnoRFnoFOOTNOTES.tex
\begin{table*}
\begin{center}
\newcommand\rot[1]{\rotatebox{90}{\scriptsize{#1}}}
\newcommand\cR[1]{\cellcolor{red!35}#1}
\newcommand\cO[1]{\cellcolor{orange!35}#1}
\newcommand\cY[1]{\cellcolor{yellow!35}#1}
\newcommand\cG[1]{\cellcolor{green!25}#1}
\setlength{\tymin}{1pt}
\setlength{\tabcolsep}{3pt}
\newcommand\acht{\scriptsize{\unit[868]{MHz}}}
\newcommand\neun{\scriptsize{\unit[915]{MHz}}}
\renewcommand{\arraystretch}{1.5}
\caption{Standard deviations of means of the measurements with \unit[-60~$\pm$~1]{dB~attenuation} and \unit[18]{m} cable length.}
\newcolumntype{Z}{>{\centering\arraybackslash}X}
\newcolumntype{R}{>{\raggedright\arraybackslash}X}
\newcolumntype{L}{>{\raggedleft\arraybackslash}X}
\begin{tabularx}{0.9\textwidth}{p{50pt}p{4pt}p{4pt}p{28pt}ZZZZZZZZZ}
	\toprule
	\multicolumn{4}{c}{Size of measurement batch} & \textbf{1} & \textbf{20} & \textbf{50} & \textbf{100} & \textbf{200} & \textbf{500} & \textbf{1000} & \textbf{2000} & \textbf{5000}\\
	\midrule[0.09 em]
	\multirow{4}{0.6in}{Standard deviations of means~[m]}                         																									                               & \multirow{4}{6pt}{\rot{\unit[250]{kb/s}}} & \multirow{2}{6pt}{\rot{GFSK}} & \acht                                                                                 & \cR{12.80} & \cR{2.85} & \cO{1.80} & \cO{1.26} & \cY{0.88} & \cY{0.58} & \cG{0.41} & \cG{0.29}& \cG{0.19} \\
	   & & & \neun                                                                                                                                                         & \cR{12.83} & \cR{2.87} & \cO{1.82} & \cO{1.28} & \cY{0.91} & \cY{0.55} & \cG{0.37} & \cG{0.24} & \cG{0.15} \\
	   & & \multirow{2}{6pt}{\rot{FSK}} & \acht                                                                                                                            & \cR{6.09} & \cO{1.37} & \cY{0.86} & \cY{0.61} & \cG{0.44} & \cG{0.27} & \cG{0.19} & \cG{0.13} & \cG{0.08} \\
	   & & & \neun                                                                                                                                                         & \cR{6.26} & \cO{1.40} & \cY{0.88} & \cY{0.62} & \cG{0.44} & \cG{0.28} & \cG{0.21} & \cG{0.14} & \cG{0.09} \\
	\cmidrule[0.03em]{2-13}
	\multicolumn{4}{c}{Measurement duration\footnotemark[1]~[ms]} & 1.4 & 28 & 70 & 140 & 280 & 700 & 1400 & 2800 & 7000\\
	\midrule[0.09 em]
	
	\multirow{4}{0.6in}{Standard deviations of means [m]}                                                                                                           & \multirow{4}{6pt}{\rot{\unit[38.4]{kb/s}}} & \multirow{2}{6pt}{\rot{GFSK}} & \acht                                                                                & \cR{78.93} & \cR{17.73} & \cR{11.19} & \cR{7.87} & \cR{5.59} & \cR{3.57} & \cR{2.59} & \cO{1.71} & \cO{1.12} \\
	   & & & \neun                                                                                                                                                         & \cR{78.36} & \cR{17.56} & \cR{11.12} & \cR{7.88} & \cR{5.62} & \cR{3.63} & \cR{2.42} & \cO{1.69} & \cO{1.04} \\
	   & & \multirow{2}{6pt}{\rot{FSK}} & \acht                                                                                                                            & \cR{30.10} & \cR{7.11} & \cR{4.50} & \cR{3.17} & \cR{2.27} & \cO{1.45} & \cO{1.02} & \cY{0.77} & \cG{0.43} \\
	   & & & \neun                                                                                                                                                         & \cR{29.85} & \cR{7.04} & \cR{4.55} & \cR{3.20} & \cR{2.25} & \cO{1.41} & \cO{1.03} & \cY{0.77} & \cG{0.49} \\
	\cmidrule[0.03em]{2-13}
	\multicolumn{4}{c}{Measurement duration~[ms]} & 4.3 & 86 & 215 & 430 & 860 & 2150 & 4300 & 8600 & 21500\\
	\midrule[0.09 em]
	
	\multirow{2}{0.6in}{Standard deviations of means~[m]}           							                                                                                & \multirow{2}{6pt}{\rot{\unit[1.2]{kb/s}}} & \multirow{2}{6pt}{\rot{GFSK}} & \acht                                                                                 & \cR{1741.33} & \cR{394.14} & \cR{253.2} & \cR{179.76} & \cR{126.22} & \cR{86.41} & \cR{64.92} & \cR{58.48} & \cR{29.16} \\
	   & & & \neun                                                                                                                                                        & \cR{1745.12} & \cR{395.28} & \cR{250.51} & \cR{177.88} & \cR{123.46} & \cR{81.49} & \cR{55.20} & \cR{33.32} & \cR{24.51} \\
	\cmidrule[0.03em]{2-13}
	\multicolumn{4}{c}{Measurement duration~[ms]} & 110 & 2200 & 5500 & 11000 & 22000 & 55000 & 110000 & 220000 & 550000\\

	\bottomrule
\end{tabularx}
\vspace*{2pt}
\renewcommand{\arraystretch}{1.0}
\begin{tabularx}{0.95\textwidth}{p{12pt}lZZZZp{10pt}}
&\rule{0pt}{1em}Error coloring: & \cR{\unit[2]{m}~$<\sigma$} & \cO{\unit[2]{m}~$\leq\sigma<$~\unit[1]{m}} & \cY{\unit[1]{m}~$\leq\sigma<$~\unit[0.5]{m}} & \cG{$\sigma<$~\unit[0.5]{m}}&\\
\end{tabularx}
\vspace*{1pt}
\renewcommand{\arraystretch}{0.1}
\begin{tabularx}{0.9\textwidth}{X}
\footnotemark[1]All values were calculated from the measured time of single measurement.\\
\end{tabularx}
\label{tab:stdev}
\end{center}
\end{table*}